\documentclass[aps,prl,twocolumn, superscriptaddress,amsmath,showpacs, showkeys, floatfix]{revtex4-1}

\usepackage[latin1]{inputenc}% ISO Latin 1 encoding
\usepackage{graphicx}% Include figure files
\usepackage{dcolumn}% Align table columns on decimal point
\usepackage{bm}% bold math
\usepackage{setspace}

\newcommand {\dg} {\ensuremath{^{\circ}}}

\newcommand {\V}[1] {$\mathbf{#1}$} % 	not for use within math mode!

\newcommand {\Tz} {\ensuremath{T_{0}}}

\newcommand {\CeRuAl} {CeRu$_{2}$Al$_{10}$}
\newcommand {\CeOsAl} {CeOs$_{2}$Al$_{10}$}
\newcommand {\CeTAl} {Ce$T_{2}$Al$_{10}$}

\begin{document}

\title{Anisotropic spin-dynamics in the Kondo semiconductor  \CeRuAl\
%with antiferromagnetic order
}

%\author{Author's name}
%\thanks{}
%\affiliation{Your affiliation}



\author{Julien Robert}
\email[e-mail address: ]{julien.robert@cea.fr}
\author{Jean-Michel Mignot}
\author{Sylvain Petit}
%\author{Gilles Andr\'{e}}
%\thanks{}
\affiliation{Laboratoire L\'{e}on Brillouin, CEA-CNRS, CEA/Saclay, 91191 Gif sur Yvette (France)}

\author{Paul Steffens}
\affiliation{Institut Laue-Langevin, BP 156, 38042 Grenoble Cedex 9, France}

\author{Takashi Nishioka}
\author{Riki Kobayashi}
\author{Masahiro Matsumura}
\affiliation{Graduate School of Integrated Arts and Science, Kochi University, Kochi 780-8520 (Japan)}

\author{Hiroshi Tanida}
\author{Daiki Tanaka}
\author{Masafumi Sera}
\affiliation{Department of Quantum Matter, ADSM, Hiroshima University, Higashi-Hiroshima, 739-8530 (Japan)}

\date{\today}

%                                                                      Abstract

\begin{abstract}
Spin dynamics in the new Kondo insulator compound \CeRuAl\ has been studied using unpolarized and polarized neutron scattering on single crystals.  In the unconventional ordered phase forming below $T_0 = 27.3$ K, two excitation branches are observed with significant intensities, the lower one of which has a gap of $4.8 \pm 0.3$ meV and a pronounced dispersion up to $\approx 8.5$ meV. Comparison with RPA magnon calculations assuming crystal-field and anisotropic exchange couplings captures major aspects of the data, but leaves unexplained discrepancies, pointing to a key role of direction-specific hybridization between 4$f$ and conduction band states in this compound.
\end{abstract}

\pacs{
71.27.+a,	% Strongly correlated electron systems; heavy fermions
%71.28.+d,	% Narrow-band systems; intermediate-valence solids (for magnetic aspects, see 75.20.Hr and 75.30.Mb in magnetic properties and materials)
%71.70.Ch,	% Crystal and ligand fields
%71.70.Ej,	% Spin-orbit coupling, Zeeman and Stark splitting, Jahn-Teller effect
%71.70.Gm,	% Exchange interactions
75.20.Hr,		% Local moment in compounds and alloys; Kondo effect, valence fluctuations, heavy fermions (see also 72.15.Qm Scattering mechanisms and Kondo effect)
%75.25.+z,	        % Spin arrangements in magnetically ordered materials (including neutron and spin-polarized electron studies, synchrotron-source X-ray scattering, etc.) (for devices exploiting spin polarized transport, see 85.75.?d)
75.30.Ds,		%Spin waves (for spin-wave resonance, see 76.50.+g)
%75.30.Et,	% Exchange and superexchange interactions (see also 71.70.?d Level splitting and interactions)
75.30.Gw,		%Magnetic anisotropy
%75.30.Kz,	%Magnetic phase boundaries (including magnetic transitions, metamagnetism, etc.)
75.30.Mb,	% Valence fluctuation, Kondo lattice, and heavy-fermion phenomena (see also 71.27.+a Strongly correlated electron systems, heavy fermions)
%75.47.Np,	% Metals and alloys
78.70.Nx	% Neutron inelastic scattering
}


\keywords{\CeRuAl, inelastic neutron scattering, Kondo insulator, singlet ground state, spin-gap, magnon, anisotropy}

\maketitle

% ///////////////////////////////////////////////////////////////////// MAIN TEXT/////////////////////////////////////////////////////////////////////

%\section{\label{sec:intro}Introduction}

The Ce$M_2$Al$_{10}$ ($M =$ Fe, Ru, Os) compounds form a new family of Ce-based intermetallic materials with fascinating, but hitherto elusive, magnetic and transport properties. Below room temperature, they show evidence of a Kondo-insulator regime, with an increase in the electrical resistivity on cooling ascribed to the opening of a narrow ``hybridization gap'' in the electronic density of states \cite{Nishioka'09, Strydom'09}. In the standard approach \cite{Risebg00}, this mechanism should ultimately lead to a nonmagnetic, many-body singlet ground state for $T \rightarrow 0$, as was observed experimentally for the vast majority of Kondo-insulator compounds known to date. In contrast, \CeRuAl\ and \CeOsAl\ order magnetically below $T_0 = 27.3$ K and 28.7 K, respectively \cite{Nishioka'09}. Their structure is antiferromagnetic (AF) with the simple wavevector $\mathbf{k}_{\mathrm{AF}} = (0,1,0)$ \cite{Robert'10,Khalyavin'10,Mignot'11}. However, there is strong experimental evidence that this ordering  cannot be explained by conventional Ruderman-Kittel Kasuya-Yosida (RKKY) exchange alone: $T_0$ seems unrealistically high in view of the large Ce--Ce interatomic distances (5.26 \AA), of the weak ordered antiferromagnetic moment ($\mu_{AF}=0.32(4)$--$0.4 \mu_B$ \cite{Mignot'11,Khalyavin'10,Kato'11} for $M=$ Ru) derived from neutron diffraction measurements, and of the much lower Néel temperatures found in other  $T$Ru$_2$Al$_{10}$ compounds ($T_N = 16.5$ K in GdRu$_2$Al$_{10}$ \cite{Kobayashi'11}). It was also reported that $T_0$ increases with the application of pressure \cite{Nishioka'09}, contrary to the general trend in  Ce Kondo compounds. This unique situation has attracted considerable interest because it seems to challenge widely accepted views on Kondo insulators.  Various interpretations have been proposed in terms of ($i$) a charge density wave associated with an energy gap opening preferentially along the $b$ direction \cite{Kimura'11.Ru, Kimura'11.Os},  ($ii$) a spin-Peierls state due to the formation of spin-singlet pairs  \cite{Tanida'10.1, Hanzawa'10.1, Hanzawa'10.2}, or  ($iii$) a resonating-valence-bond state \cite{Hanzawa'11a}. Quite remarkably, despite the large anisotropy of the paramagnetic susceptibility with $\chi_a \gg \chi_c \gg \chi_b$, the ordered AF moments align along the $c$ direction \cite{Khalyavin'10, Mignot'11}. In Refs.~\cite{Kondo'11,Tanida'12}, this discrepancy was suggested to arise from conduction-electron--$f$-electron ($c$--$f$) hybridization occurring predominantly along $a$, and suppressing $\chi_a$ accordingly through the formation of a (Kondo) spin singlet. A detailed study of the spin dynamics is of primary importance to sort out this problem. Inelastic neutron scattering (INS) experiments performed previously on powder samples have evidenced the opening of a large spin-gap in the ordered state, with a broad excitation centered at $\Delta_{\mathrm{SG}} = 8$ meV and 11 meV in \CeRuAl\ \cite{Robert'10} and \CeOsAl\ \cite{Adroja'10}, respectively. However, mode dispersion and anisotropy were obscured by powder averaging, and it could not be decided whether the observed magnetic signal arose from dispersive magnon branches with an anisotropy gap or, e.g., from singlet-triplet transitions with sizable dispersion and/or damping. The possibility of a lattice contribution could also not be ruled out. In this Letter, we report unpolarized and polarized INS experiments performed on single-crystal \CeRuAl. The spectra reveal well-defined dispersive excitations with a gap of 4.8 meV at the AF zone center. They exhibit a remarkable anisotropy which does not correspond to a standard precession of spin wave modes.
Overall agreement  with the experimental results can be achieved phenomenologically in a RPA model by assuming a strongly anisotropic bilinear exchange interaction ${\cal J}^{c} \gg {\cal J}^{a}, {\cal J}^{b}$. However, remaining inconsistencies are thought to reflect anisotropic hybridization effects, whose role was  suspected from previous studies \cite{Tanida'12}.

Thirteen single crystals of \CeRuAl\ (orthorhombic, $Cmcm$ space group, No. 63) with dimensions comprised between 1 and 4 mm, for a total mass of about 500 mg, were grown by an Al-flux method, and co-aligned with their $b$ axes vertical on an Al sample holder.
%\footnote{One piece of $m=29$ g was removed after the 2T experiment.}
 An effective mosaicity of about 3 degrees was estimated from the neutron rocking curves, which was sufficient for the present experiment. Excitation spectra were measured in the $(a^*, c^*)$ scattering plane, first using unpolarized neutrons on the 2T triple-axis spectrometer at LLB-Orphée (Saclay), then with linear polarization analysis on IN20 at the ILL (Grenoble). Finally, the crystals were reoriented with the $c$ axis vertical on a lighter sample holder, and measured with unpolarized neutrons on IN8 (ILL) in a 6 T cryomagnet. Spectra were recorded at fixed final energy, $E_f=14.7$ meV, using a pyrolytic graphite, PG002 (2T) or Si111 (IN8) monochromator and a PG002 analyzer, with a PG filter placed on the scattered beam, or (polarized neutrons on IN20) a Heussler monochromator and analyzer.

%Figure Escans_unpol
 \begin{figure}  [t] %[h] % [h]ere, [t]op, [b]ottom, [p]age of floats
 	\centering
	\includegraphics [clip, width=0.55\columnwidth, angle=-90] {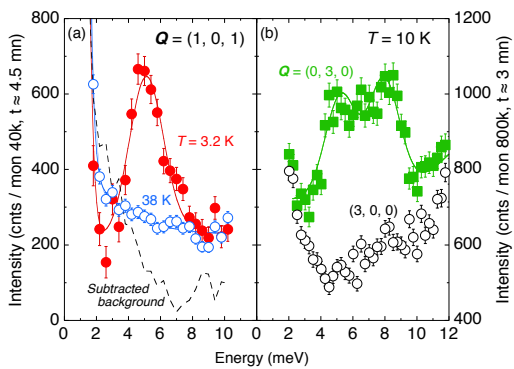}
	\caption{\label{Escans_unpol} (Color online) Energy scans measured on 2T (a) and IN8 (b) at $k_f = 2.662$ \AA$^{-1}$ for different AF zone centers: (a) $\mathbf{Q} = (1, 0, 1)$ at $T = 3.2$ K (closed circles) and 38 K (open circles). A steep, temperature independent, background (dashed line) was estimated by assuming the magnetic signal at 38 K to be \V{Q} independent (verified for other \V{Q} vectors), then subtracted from  the measured data. (b)  $\mathbf{Q} = (3, 0, 0)$ (open circles) and (0, 3, 0) (closed squares) at $T = 10$ K. In (a) and (b), solid lines represent fits to the  data using Gaussian (inelastic) and Lorentzian (quasielastic) line shapes.}
\end{figure}

%Figure dispersion
 \begin{figure}  [t] %[h] % [h]ere, [t]op, [b]ottom, [p]age of floats
 	\centering
	\includegraphics [trim= 0 0mm 0 0, clip, width=0.85\columnwidth, angle=0] {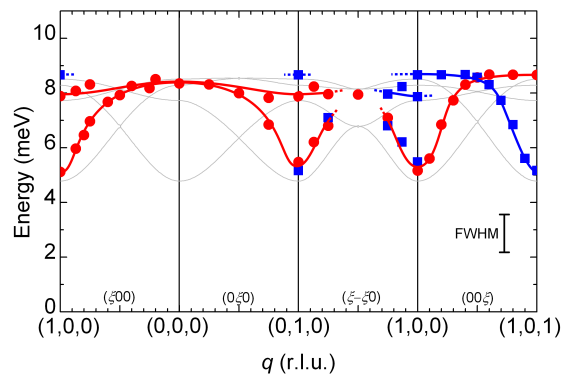}
	\caption{\label{dispersion} (Color online) Dispersion of the magnetic excitations in \CeRuAl. Closed circles: results from the unpolarized neutron experiments; red and blue symbols: guides to the eye; grey lines: RPA calculations (see text).}
\end{figure}

%Figure maps
 \begin{figure}  [t] %[h] % [h]ere, [t]op, [b]ottom, [p]age of floats
 	\centering
	\includegraphics [clip, width=0.75\columnwidth, angle=0] {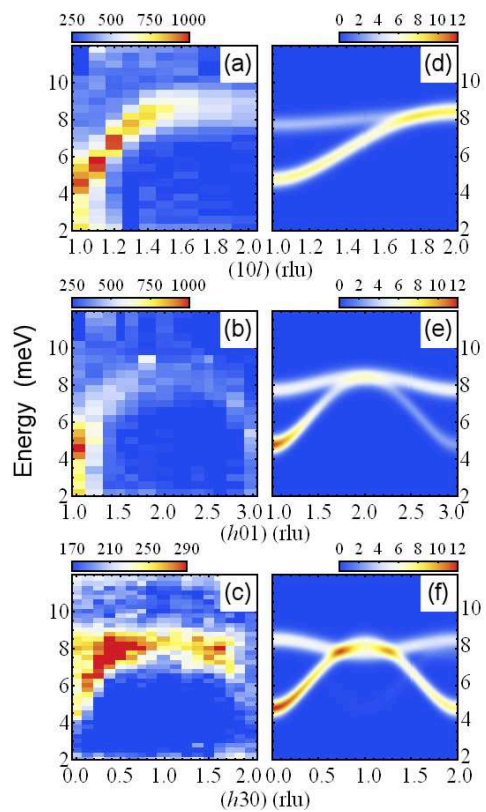}
	\caption{\label{maps} (Color online) Left: intensity maps derived from the spectra measured on (a, b): 2T at $T = 3.2$ K and (c): IN8 at $T = 10$ K  for three directions in reciprocal space. Different color scales are used to reflect the different counting rates on the two spectrometers. Right: RPA calculations (see text).}
\end{figure}

Constant-\V{Q} scans have been performed using unpolarized neutrons for momentum transfers lying in the ($a^*$, $c^*$) (2T) and  ($a^*$, $b^*$) (IN8) planes. Representative spectra are presented in Fig.~\ref{Escans_unpol}. For $T = 3.2$ K (2T) or 10 K (IN8), one or two distinct modes are visible depending on the \V{Q} vector. The dispersion is significant, with a gap of $4.8 \pm 0.3$ meV at the AF zone centers (Fig.~\ref{dispersion}). Near the zone boundary, the excitations reach $8.5 \pm 0.3$ meV, with a flat region corresponding to the peak observed just above 8 meV in the previous powder experiments \cite{Robert'10}. Intensity maps for three particular directions, $\mathbf{Q} = (1, 0, l)$, $(h, 0, 1)$ and $(h, 3, 0)$, are presented in Fig.~\ref{maps}. The existence of (at least) two modes is best evidenced in scans at $\mathbf{q} = \mathbf{k}_{\mathrm{AF}}$, e.g. for $\mathbf{Q} = \boldsymbol{\tau}_{020} + \mathbf{k}_{\mathrm{AF}} = (0, 3, 0)$. On the other hand, the lower branch shows no detectable intensity at the AF \V{Q} vector (1, 0, 2) (of the form $(h,0,l)$ with $l$ even), as  shown in Fig.~\ref{maps}(a). Another important observation is that the magnetic intensity of the lower branch is strongly suppressed for scattering vectors whose orientation is close to the $a^*$ axis, such as \V{Q} = (3, 0, 0) as compared to (0, 3, 0) (Fig.~\ref{Escans_unpol}). This suggests that dynamical correlations $\langle m_i^b m_j^b \rangle$ and $\langle m_i^c m_j^c \rangle$ between moment components perpendicular to the $a$ axis are weak, and $\langle m_i^a m_j^a \rangle$ correlations dominate the magnetic response. When temperature increases to $T = 38$ K $> T_0$, the inelastic magnetic peak at $4.8 \pm 0.3$ meV is suppressed and replaced by a sloping intensity at low energy. The latter signal shows no pronounced \V{Q} dependence [apart from the appearance of a strong extra background near \V{Q} = (1, 0, 1), see Fig.~\ref{Escans_unpol}(a)], and is thus ascribed to quasielastic (QE) fluctuations. Spectra along $(h, 3, 0)$, $(0, 2+k, 0)$, and $(h, 3-h, 0)$ were also measured at the base temperature in an applied field ($H \parallel c$) of 5 T, above the moment reorientation transition from $\mathbf{m}_{\mathrm{AF}} \parallel c$ to $\mathbf{m}_{\mathrm{AF}} \parallel b$, known to occur at $H^{*} \approx 4$ T \cite {Tanida'10.3,Tanida'11}. No sizable change was observed with respect to the $H =0$ data.

%Figure Escans_pol and maps_pol_

 \begin{figure}  [t] %[h] % [h]ere, [t]op, [b]ottom, [p]age of floats
 	\centering
	\includegraphics [width=0.55\columnwidth, angle=-90] {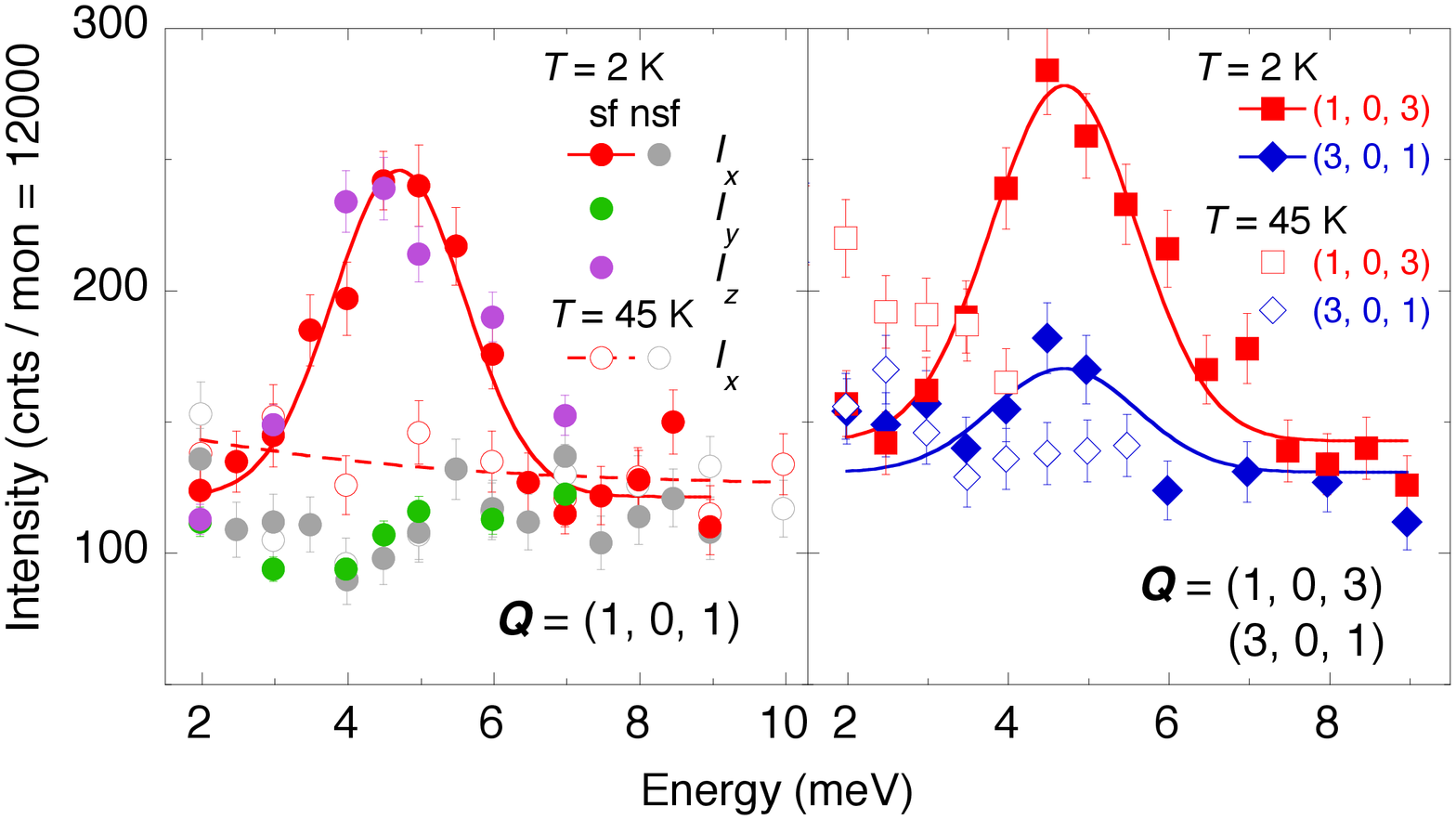}\\
	\vspace {4pt}
	\includegraphics [trim = -20mm 0 0 6mm, clip, width=0.90\columnwidth, angle=0] {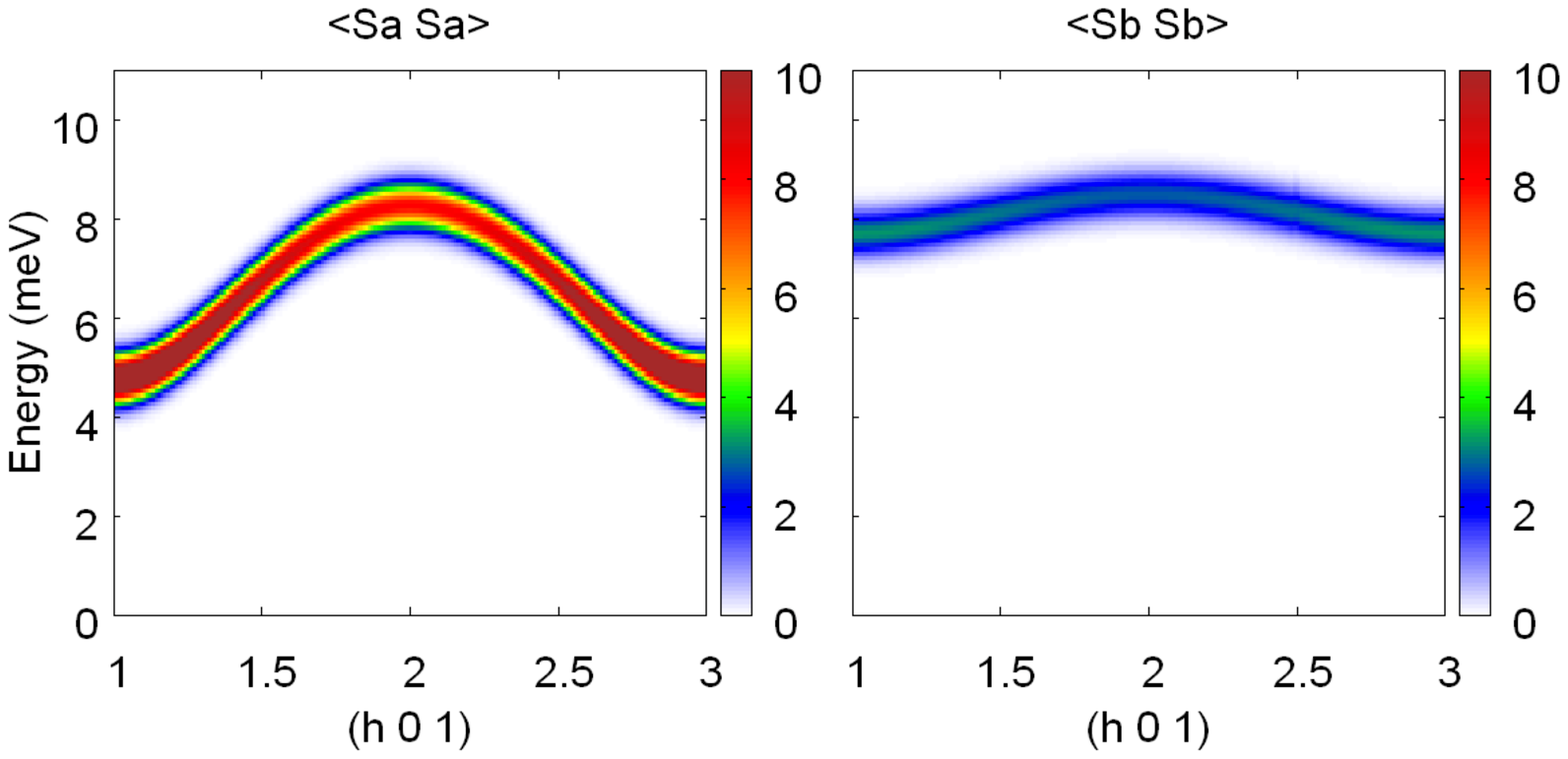}

	\caption{\label{Escans_pol} (Color online) Polarization analysis  on IN20 ($k_f = 2.662$ \AA$^{-1}$). Upper frames : energy scans measured at $T= 2$ and 45 K for different AF zone centers. Left: $\mathbf{Q} = (1, 0, 1)$; grey symbols denote the NSF intensity measured with $\mathbf{P_0} \parallel x$, and other symbols SF intensities for $\mathbf{P_0} \parallel x$, $y$ or $z$. Right: $\mathbf{Q} = (1, 0, 3)$ and $(3, 0, 1)$ (red squares and blue diamonds, respectively); the plot shows SF intensities for $\mathbf{P_0} \parallel x$. Full (dashed) lines represent intensities calculated using Gaussian (Lorentzian) spectral functions. Lower frames:  intensity maps along the $(h, 0, 1)$ direction for the two transverse components $\langle m_i^a m_j^a \rangle$ (left) and  $\langle m_i^b m_j^b \rangle$ (right) of the magnetic correlations using the exchange parameters listed in Table~\ref{param}.} 
\end{figure}

Neutron polarization analysis provides further insight into the anisotropy of the magnetic response. Fig.~\ref{Escans_pol} (upper frames) shows intensities measured in the spin-flip (SF) and non-spin-flip (NSF) channels at different scattering vectors. Let us first consider the results for $\mathbf{Q} = (1, 0, 1)$. One sees that the NSF signal (measured with the incident polarization $\mathbf{P_0} \parallel \mathbf{Q}$) is featureless and temperature independent, confirming the magnetic origin of both the 4.8-meV peak below 10 K and the QE signal at $T > T_0$  found in the unpolarized neutron experiments. The SF intensities measured for different directions of incident polarization $\mathbf{P_0} \parallel x, y$ or $z$
\footnote{The standard notation is used, in which $x$ is oriented along \V{Q}, $z$ normal to the scattering plane, and  $\{x, y, z\}$ forms a direct coordinate system.}
are found to fulfill $I_x \approx I_z$ and $I_y \approx 0$. Using the standard expressions  \cite{Regnault'01}

\begin{subequations}
\label{eqn1_whole}
\begin{align}
\label{eqn1a}
I_x^{\mathrm{sf}} & \propto M_{aa}(\mathbf{q})\sin^2(\alpha) + M_{bb}(\mathbf{q}) + M_{cc}(\mathbf{q})\cos^2(\alpha)] \\
\label{eqn1b}
I_y^{\mathrm{sf}} & \propto M_{bb}(\mathbf{q}) \\
\label{eqn1c}
I_z^{\mathrm{sf}} & \propto M_{aa}(\mathbf{q})\sin^2(\alpha) + M_{cc}(\mathbf{q})\cos^2(\alpha),
\end{align}
\end{subequations}

\noindent where $M_{pp}$ is the dynamic structure factor associated with pair correlations of the moment component $m_p$ ($p=\{a, b, c\}$), and $\alpha$ the angle between \V{Q} and the $a^*$ axis, one comes to the conclusion that correlations of the $m_b$ components must vanish to the precision of the present measurement. For $\mathbf{Q} = (1, 0, 1)$, $\alpha$ is very close to 45\dg\ since the lattice parameters $a$ and $c$ are nearly equal. In contrast, the scattering vectors $\mathbf{Q} = (1, 0, 3)$ and (3, 0, 1) correspond to the same reduced \V{q} vector (AF zone center) and nearly equal values of the dipole magnetic form factor, but their $\alpha$ angles are quite different (71.6\dg\ and 18.4\dg, respectively). From Fig.~\ref{Escans_pol}, the ratio of the magnetic excitation intensities $I_x^{\mathrm{sf}}$ for those two spectra is about 3.25, which implies that correlations of $a$ components dominate. Assuming $M_{bb}$ to be strictly zero, and solving Eqns.~\ref{eqn1a} and \ref{eqn1c}, one gets $M_{aa}/M_{cc} \approx 5$. In a magnon picture, such a difference can be understood by noting that $M_{aa}$ and $M_{cc}$ correspond, respectively, to transverse and longitudinal excitation modes of the AF magnetic structure. On the other hand, the strong difference between the transverse components along $a$ and $b$ is quite remarkable and requires a very unusual anisotropy to exist in this material.

To analyze this magnetic response, we have performed calculations assuming bilinear exchange interactions, 
${\cal H}_{i,j} = \sum_{\alpha} {\cal J}^{\alpha}S_i^{\alpha}S_j^{\alpha}$,  between near-neighbor $(i, j)$ Ce sites. Both a standard spin wave model, and random-phase approximation (RPA) calculations were investigated. In the following, we will focus on the second approach, which can treat anisotropy effects in a more realistic way. The crystal-field (CF) parameters for the Ce$^{3+}$ $J = 5/2$ ground state, $(B_2^0, B_2^2, B_4^0, B_4^2, B_4^4)  = (-1.326, -29.236, +1.013,$$ -1.747, -5.317)$ K, choosing $c$ as the quantization axis,  were taken from Strigari's work \cite{Strigari'12}, and correspond to a sequence of three doublets at 0, 354 K, and 535 K. The resulting single-ion anisotropy has an easy $a$ axis, as required by the magnetic susceptibility measured in the paramagnetic regime. Therefore, in this simple picture, one has to assume that ${\cal J}^{c}$ is much larger than ${\cal J}^{a}$ and ${\cal J}^{b}$ to ensure that the AF ordered moments properly align along the $c$ axis. 

%:::::::::::::::::::::::::::::::::::::::::::::::::::   Table I   ::::::::::::::::::::::::::::::::::::::::::::::::::::::::::::::
\begin{table}
\centering
\caption{\label{param}Anisotropic exchange parameters (in units of K) used in the RPA calculation.
Atomic positions $(x_i, y_i, z_i)$,  $i = 1: (0, y, \frac{1}{4}); 2: (1/2,  \frac{1}{2} + y,  \frac{1}{4}); 3: (1/2,  \frac{1}{2}-y,  \frac{3}{4}), 4: (0, -y,  \frac{3}{4})$, with $y = 1.1239(3)$ \cite{Moriyoshi'13}
}
\begin{ruledtabular}
\begin{tabular}{c c c c c }
Ce pairs $(i, j)$ & ${\cal J}^{a}$ & ${\cal J}^{b}$ & ${\cal J}^{c}$ \\
(1,4); (2,3) & 2.7 & 2.7 & 58\\    
(1,3); (2,4) & -0.9 & -0.9 & -0.9\\ .
(1,2); (3,4) & 1.1 & 1.1 & 1.1\\ 
\end{tabular}
\end{ruledtabular}
\end{table}
%:::::::::::::::::::::::::::::::::::::::::::::::::::::::::::::::::::::::::::::::::::::::::::::::::::::::::::::::::::::::::::::::::::

Fair overall agreement can be obtained between the calculations and the experimental excitation spectra below \Tz, using the set of exchange constants listed in Table~\ref{param}. As can be seen in Fig.~\ref{maps}, the observation of two branches with significant spectral weight (from a total of 4), as well as the general $Q$ dependence of their intensities along different symmetry directions, or the anisotropy of the correlations (Fig.~\ref{Escans_pol}, lower frames) can be accounted for. Furthermore, salient features of the experimental data are well reproduced in the calculations, such as the vanishing of the lower dispersive branch near the (1,0,2) AF zone center (upper frames in Fig.~\ref{maps}), or the significant intensity exhibited by the upper branch near $\mathbf{Q} = (0, 3, 0)$ [lower frames, in accordance with Fig.~\ref{Escans_unpol} (b)], in contrast to, e.g., $\mathbf{Q} = (1, 0, 1)$ [upper frames and Fig.~\ref{Escans_unpol}(a)]. 
On the other hand, notable quantitative differences exist:  the initial slopes of the dispersions are much steeper than predicted by the calculation, and the calculated energy of the upper mode is too high. We believe that this discrepancy results from the unrealistically large ${\cal J}^c$ value required to keep the ordered moments aligned along the $c$ axis despite the strong single-ion anisotropy favoring the $a$ axis. Simulations done in the simpler Holstein-Primakoff spin-wave approximation indeed showed that the agreement improves if one reduces this single-ion anisotropy and, correspondingly, the anisotropic component of the exchange tensor. Recent simulations performed in a mean-field, two-sublattice, model \cite{Kunimori'13} further indicate that anisotropic exchange parameters large enough to overcome the single-ion $a$-axis anisotropy inevitably result in a large ordered moment, contrary to the experimental observation of $\mu_{AF} = 0.32(4)$--$0.42(1) \mu_B$. This could raises the question of whether the CF model of Ref.~\cite{Strigari'12} used in the present calculations overestimate the single-ion anisotropy. 
Meanwhile, there is growing experimental evidence, as discussed in recent papers \cite{Kimura'11.Ru, Tanida'12}, that direction-selective hybridization of $4f$ orbitals with conduction band states plays a key role in the peculiar magnetism of the \CeTAl\ compounds. 
This has been proposed to explain the anomalous magnitude of the single-ion anisotropy in \CeRuAl, as compared to that of Nd$T_2$Al$_{10}$, as well as the lack of a sizable anomaly ($\Delta l / l < 10^{-6}$) in the longitudinal magnetostriction at the critical field $H^{*}_{\parallel c} \approx 4$ T where the AF moment direction reorients from  $c$ to  $b$  \cite{Tanida'12} (possibly related to the intriguing lack of field dependence of the magnetic excitation spectra found in the present measurements). It has been argued \cite{Moriyoshi'13} that, owing to specifics of the YbFe$_2$Al$_{10}$-type crystal structure, $f$--$p$ hybridization takes place predominantly within the $(a,c)$ plane, especially with the Al(2) atoms located in the $a$ direction with respect to the Ce site. This hybridization could result in a suppression of the magnetic components along $a$, thereby favoring the alignment of the ordered AF moments along $c$. Such a picture provides an appealing physical basis for the reduction of the single-ion anisotropy hypothesized in the above discussion. In the case of \CeOsAl, it has been argued \cite{Kondo'11} that the gap in the magnetic excitation spectrum, associated with the formation of a singlet state, starts to develop below the temperature of the maximum in the magnetic susceptibility $\chi_a (T)$, well above the onset of the AF order. Such effects are clearly beyond the scope of the  simple magnon model presented above, which basically treats the spin gap as an anisotropy gap, but should  be included in a more realistic treatment.

In conclusion, the present study provides detailed insight into the spin dynamics of \CeRuAl, and emphasizes the most peculiar anisotropy of the magnetic correlations occurring in the AF ordered state. The results could be partly accounted for using a magnon model treated in the RPA approximation. However quantitative discrepancies suggest that this picture should be regarded as phenomenological, and support the idea that anisotropic $c$--$f$ hybridization plays a key role in this material. Proper theoretical consideration of such effects should open the way to a unifying view of static and dynamic aspects of magnetism in this family of compounds.


\begin{acknowledgments}
We thank P. Baroni and F. Maignen for technical support, and E. Wheeler and A. S. Ivanov for help during the work at the ILL.
\end{acknowledgments}

% ///////////////////////////////////////////////////////////////// END MAIN TEXT //////////////////////////////////////////////////////////////

%\bibliography{CeRu2Al10}           % in case an external BibTeX library is used

%

\end{document}